\documentclass[aps,twocolumn,showpacs]{revtex4}

\begin{document}

\title{The Scaling of Maximum and Basal Metabolic Rates of Mammals and Birds}

\author{Lauro A. Barbosa$^{\ast}$, Guilherme J. M. Garcia$^{\dagger}$ and 
Jafferson K. L. da Silva$^{\ddagger}$}
\affiliation{
Departamento de F\'{\i}sica, Instituto de Ci\^encias Exatas,
Universidade Federal de Minas Gerais, C. P. 702, CEP 30123-970, Belo
Horizonte, MG - Brazil}

\date{\today}\widetext


\begin{abstract}
Allometric scaling is one of the most pervasive laws in biology. Its origin,
however, is still a matter of dispute. Recent studies have established that 
maximum metabolic rate scales with an exponent larger than that found for 
basal metabolism. This unpredicted result sets a challenge that can decide 
which of the concurrent hypotheses is the correct theory. Here we show that
both scaling laws can be deduced from a single network model. Besides the 
3/4-law for basal metabolism, the model predicts that maximum metabolic rate
scales as $M^{6/7}$, maximum heart rate as $M^{-1/7}$, and muscular capillary
density as $M^{-1/7}$, in agreement with data.
\end{abstract}

\maketitle

Metabolic rate $B$ and body mass $M$ are connected by the scaling relation $B=aM^b$, where $a$ is a 
constant and $b$ is the allometric exponent. Kleiber's law \cite{ref1} ($b=3/4$) characterizes 
the basal metabolism of almost all organisms \cite{ref2,ref3,ref4}, including mammals and birds. 
The predominance of quarter-power scaling in biology is experimentally evident \cite{ref2,ref3,ref4} and
is supported by recent theories based on the design of resource distribution networks \cite{ref5,ref6}. 
Maximum metabolic rate, however, scales with an exponent somewhat larger than $3/4$ ($b\approx 0.86$) 
\cite{ref4,ref7,ref8,ref9,ref10,ref11}.
This result raises the question whether the scaling of maximal metabolic rate is governed by 
mechanisms different from those determining basal metabolism \cite{ref4,ref8,ref9}. 
Here we reformulate the ideas of West, Brown and Enquist \cite{ref5} to show that {\sl both} scaling
laws can be deduced from a {\sl single} network model.
Although it was recently demonstrated that the exponent $3/4$ cannot be consistently derived from
the assumptions of West et al. \cite{ref5,ref12}, we show that, using a different set of assumptions
and the appropriate impedances, the model correctly predicts the $3/4$-law for basal metabolism.

The basal and maximum metabolic rates set the limits of the energetic range of endothermic animals.
Considerable effort has recently been invested to understand the scaling of these variables 
\cite{ref4,ref5,ref6,ref7,ref8,ref9,ref10,ref11,ref12,ref13,ref14,ref15,ref16}.
In particular, maximal performance is very informative about animal design since during strenuous
exercise the circulatory and respiratory systems are stressed to their uttermost. 
Recent studies have established that maximal metabolic rate (MMR) scales with an exponent larger 
than that found for basal metabolism, suggesting that the scaling of MMR is determined by 
mechanisms different from those controlling basal metabolic rate (BMR) \cite{ref4,ref8,ref9}. 
We show that this difference in the allometric exponent is a consequence of dynamic adaptations 
in the resource supply network, which are needed to cope with the distinct demands for oxygen and nutrients. 
Simple dimensional analysis linking BMR to heat loss through the surface (claimed to support a 2/3-law
for basal metabolism) \cite{ref12} and simple geometric arguments \cite{ref6} are unable to explain why different 
metabolic states have different allometric exponents.

West, Brown and Enquist (WBE) proposed that the ubiquity of the $3/4$-law lies in the
resource distribution network common to all organisms \cite{ref5}.
In order to estimate the exponent $b$ they make three assumptions:
(i) the network branchs hierarchically to supply all parts of the organism;
(ii) the terminal units, such as the capillares of the circulatory system, are body size independent;
(iii) natural selection has optimized the network so that the energy required to distribute resources
is minimal, or, equivalently, the total hydrodynamic resistance (impedance) of the network is minimal.
 From these assumptions they conclude that 
(a) the network is self-similar (fractal);
(b) the network has area-preserving branching in the first generations;
(c) blood volume $V_b$ is proportional to M;
and
(d) metabolic rate scales with $b=3/4$. 
Some matematical errors, however, have been found in the derivation of these results \cite{ref12}.
It turns out that none of these conclusions can be derived from the above hypotheses.
Nevertheless, the ideas of WBE are simple and elegant, and will be shown here to predict
correctly the allometric exponents of BMR and MMR when a different set of assumptions is adopted
and the correct impedances are used.

The impedance is not minimal for pulsatile flow in an elastic tube, nor for an 
area-preserving branching in a rigid tube. We therefore reformulate the third hypothesis
to state that 
(iii) the total impedance is minimal with respect to the radius, length, thickness and elastic modulus of the vessels. 
In other words, the impedance is minimal with respect to all variables, except the branching ratio $n$. 
In addition, we assume that 
(iv) blood volume is proportional to $M$; 
(v) the network is fractal; and
(vi) the mass of the arteries is constant for a given $M$
or, alternatively,
(vi) the total elastic energy stored in the vessels is proportional to $M$.
Hypotheses (iv) and (v) are supported empirically \cite{ref2,ref3,ref12,ref17});
assumption (vi) is discussed below.

Let us consider some aspects of the cardiovascular system \cite{ref18} and their relation
to the model. For clarity, we use the same notation as West et al. \cite{ref5}.
The circulatory network consists of elastic vessels: the aorta, large arteries, muscular arteries, arterioles 
and capillaries. Each  vessel is described by a label $k$ related to the branching level, with $k=0$ corresponding
to the aorta and $k=N$ to the capillaries. At each level, each vessel branches into $n_k$ smaller vessels,
so that the total number of vessels $N_k$ at level $k$ is $N_k=n_0 n_1\ldots n_{k-1}$.
In particular, the number of capillaries is $N_c=N_N=n_0n_1\ldots n_{N-1}$.
After branching, a vessel has a smaller length ($l_{k+1}<l_k$) and a smaller radius 
($r_{k+1}<r_k$) and the network can be characterized by the ratios $\gamma_k=l_{k+1}/l_k$ and
$\beta_k=r_{k+1}/r_k$. The heart is taken into account by considering
an oscillatory pressure $p(z,t)=p_0\exp [i(\omega t - Kz)]$, where $p_0$ is the amplitude of the pulse, $t$ is time, $z$ is
the distance along the vessel, $K=K_1+iK_2$ is the complex wavenumber and $\omega =2\pi f$ is the angular frequency. 
The velocity of the pressure pulse is given by $c=\omega /K_1=\lambda f$, with $\lambda$ the wavelength 
and $f$ the frequency; $K_2$ describes the damping of the pulse.

The thickness $h_k$ and the Young's modulus $E_k$ of the vessels also depend on the level $k$.
The vessel wall consists, in general, of three layers: the {\sl intima}, a monolayer of endothelial cells,
the {\sl media}, consisting of muscle layers, and the {\sl adventitia}, which is composed of connective tissue.
The elastic properties of the vessels vary principally due to the variable constitution of 
the media. Large arteries are distentensible and have a smaller $E$ than muscular ones. 
   Arterioles have a relatively large thickness and small changes in muscular contraction can induce 
radius changes, thus controlling flow to organs. The capillaries have no adventitia or media. 
Their function is the exchange between blood and cells.

The circulatory system of endothermic animals is a dynamical network which is adjusted
according to the metabolic state. In humans, only 20-25\% of the muscular capillaries
are perfused during rest, while all are perfused during exercise. 
This increase is due to the opening of terminal arterioles and to the rise in perfusion 
pressure at the capillary level. (Note that this implies that muscular capillary density 
is linked to MMR, instead of BMR.)   
 At rest blood flow to skeletal muscles is only a small fraction of the total (15\% in humans),
but during strenuous activities it increases to a large fraction (over $90$\% in mammals) \cite{ref8}).
 Circulating blood volume is also larger during MMR, since the veins function as
a reservoir during BMR.

In short, the transition from resting to maximum activity can be described as follows:
(a) the heart increases its rate and output;
(b) the mean arterial pressure and peripheral extramuscular resistance increase due to the sympathetic vasoconstriction;
(c) circulating blood volume increases due to vasoconstriction of the veins;
(d) extramuscular flow remains essentially constant, somewhat reduced in some organs but increased in others; and 
(e) total flow and muscular flow increase, with all muscular capillaries activated.  
The increase in heart rate and in circulating blood volume during maximal performance
highlights the elastic nature of the vessels and leads to distinct allometric exponents for BMR and MMR.

The exponent $b$ of the metabolic rate for a network with $n_k=n$ is obtained as follows. 
The volume rate of flow in a tube is
${\dot Q_k}=\pi r_k^2 u_k$, where $u_k$ is the average velocity. 
Since the fluid is conserved we have
\begin{equation}
{\dot Q_0}=N_k{\dot Q_k}=N_c\pi r_c^2u_c~~.
\end{equation}
Because the quantities related to the capillaries ($r_c$, $l_c$, and $u_c$) are size invariant (second hypothesis),
we obtain ${\dot Q_0}\sim N_c$. The fluid tranports
all the nutrients and oxygen for the metabolism, so it is natural to suppose that $B\sim {\dot Q_0}$.
Therefore we have that $B\sim N_c\sim n^N\sim M^b$. From hypothesis (iv), we know that $M$ is related to
the network quantities through the total blood volume $V_b$. Evaluating
$V_b$ explicitly in terms of $\gamma$, $\beta$ and capillary quantities yields
\begin{equation}
V_b=\sum_{k=0}^NN_k V_k=\sum_{k=0}^Nn^k\pi r_k^2l_k\sim (\gamma\beta^2)^{-N}~~,
\end{equation}
and we obtain
\begin{equation}
b=-\frac{\ln (n)}{\ln (\gamma\beta^2)}~~.\label{eq2}
\end{equation}
In order to evaluate the exponent $b$ we need $\gamma$ and $\beta$.

Let us now consider a network characterized by the variables $\{l_k,r_k,n_k,h_k,E_k\}$. The first
hypothesis can be described by $N_kl_k^3=N_{k+1}l_{k+1}^3$, the so called space-filling condition \cite{ref5},
which fixes the length fraction as $\gamma_k=n_k^{-1/3}$. The ratio $\beta_k$ comes from minimization
of the total impedance $Z$ (third hypothesis). During maximum performance, the elastic nature of the
vessels is important because circulating blood volume and volume flow take their largest values.
Moreover, heart rate is maximum and the wavelength $\lambda$ of the pressure pulse is
minimum, so that we must evaluate a {\sl local} impedance. 
Analyzing the oscillatory motion of a viscous liquid in a thin-walled  elastic tube, 
Morgan and Kiely \cite{ref19} and Womersley \cite{ref20} showed that the local impedance $Z_k$, defined by
\begin{equation}
{\dot Q(z,t)}=\frac{p(z,t)}{Z_k}=\frac{\pi r_k^2 K}{\omega \rho_0}p(z,t)~~,
\end{equation}
where ${\dot Q(z,t)}$ and $p(z,t)$ are the volume flow and the pressure wave at position $z$ and time $t$,
is given by
\begin{equation}
Z_k= \frac { AE_k^{1/2}h_k^{1/2}}  {r_k^{5/2}}~~.
\end{equation}
Here $A=\rho_0^{1/2}/\sqrt{2}\pi$ with $\rho_0$ the blood density. 
This expression was obtained by solving the Navier-Stokes equation for the liquid
coupled to the Navier equation for the vessel wall, in the linearized long wave approximation for large $r$.
Under these conditions, valid for large elastic arteries, the pressure wave 
is not attenuated ($K_2=0$), propagating with velocity $c=\omega /K_1=(E_kh_k/2\rho_0r_k)^{1/2}$.

We want to minimize the power dissipated in the circuit (cardiac output) $W=Z {\dot Q_0}^2$.
Since $B\sim {\dot Q_0}$, this is equivalent to minimizing $Z$.
Consider the total impedance, $Z=\sum_{k=0}^N Z_k/N_k$, for a network with $n_0,n_1,\ldots ,n_{N-1}$ fixed.
For an organism of given mass $M$, we have to minimize $Z$, with a given blood volume $V_b$, a given mass
of thin-walled arteries ($M_a=\sum_k N_k2\pi r_kl_kh_k\rho_a$, $\rho_a$ is the mass density of the arteries)  and subject to 
the space-filling geometry. Defining the  Lagrange multipliers ($\lambda_b$,$\lambda_k$,$\lambda_a$),
we minimize the auxiliary function:
\begin{eqnarray}
F(r_k,l_k,h_k)&=&\sum_{k=0}^N \frac {AE_k^{1/2}h_{k}^{1/2}} {N_kr_k^{5/2}}+\lambda_b\sum_{k=0}^NN_kr_k^2l_k+\nonumber \\
              &+& \sum_{k=0}^N \lambda_kN_kl_k^3+\lambda_a\sum_{k=0}^N N_kr_kl_kh_k~~.
\end{eqnarray}
Here we have incorporated the constants in the Lagrange multipliers, and supposed that all vessels have the
same mass density. To continue let us suppose that the Young's modulus is the same for all vessels ($E_k=E$).
The requirements $\partial F/\partial l_j=\partial F/\partial r_j=\partial F/\partial h_j=0$ straighforwardly
lead to $\beta_j=n_j^{-5/12}$ and $h_j/r_j=constant$. 
Using $n_j=n$ (hypothesis (v)), we obtain from Eq. (3) that the exponent for the maximum metabolic rate is $b=6/7\approx 0.86$.
This result, larger than the basal value, is in very good agreement with experiment (see Table~\ref{tab1}). 
Additionally, we predict that capillary density scales as $N_c/M\sim M^{-1/7}$
and maximum heart rate as $B/M\sim M^{-1/7}$, which is also in good agreement with data (see Table ~\ref{tab1}).

\begin{table}
\caption{Allometric exponent $b$ describing the dependence of a variable $Y$ on body mass M 
($Y\sim M^b$).}
\label{tab1}
\begin{center}
\begin{tabular}{|l|c|c|c|}
\multicolumn{1}{c} {\bf Variable} &  \multicolumn{2}{c} {\bf Exponent} \\
\hline\hline
                           &     Predicted      &      Observed       & Ref. \\
\hline
                           &                    &                     &    \\
Basal metabolic rate       & $3/4=0.75$         &  $0.737\pm 0.026$   &  4 \\
\hline
Maximum metabolic rate     & $6/7\approx 0.86$  &  $0.828\pm 0.070$   &  4 \\
                           &                    &  $0.88\pm 0.02$     &  7 \\
                           &                    &  $0.872\pm 0.029$   &  8 \\
                           &                    &  $0.86$             & 22 \\
                           &                    &  $0.85$             & 23 \\
\hline                           
Muscular capillary density & $-1/7\approx 0.14$ &  $-0.07$ to $-0.21$ & 24 \\
\hline
Heart rate                 & $-1/7\approx 0.14$ &  $-0.17\pm 0.02$    & 25 \\
                           &                    &  $-0.16\pm 0.02$    & 26 \\
                           &                    &  $-0.15$            & 27 \\
\hline
\end{tabular}
\end{center}
\end{table}

Until now we have considered only the large arteries. We must now consider the capillaries. Morgan and Kiely \cite{ref19} 
solved the Navier-Stokes equation for the liquid coupled to the Navier equation for the vessel wall in
the limit of small radius. They found a highly attenued pressure wave with very
small velocity $c\sim (E_kh_kr_k)^{1/2}/(\omega \eta)^{1/2}$, where $\eta$ is the viscosity.
Using their results, we obtain the impedance
\begin{equation}
Z_k= \frac{ C (\eta E_kh_k)^{1/2} }{ \omega^{1/2}r_k^{7/2} }~~,
\end{equation}
where $C=8/(\sqrt{5}\pi)$. Womersley's analysis \cite{ref20} leads to a similar result.
We may then minimize the total impedance subject to the same restrictions as before, 
and within the same approximation ($E_k=E$). We readily obtain $\beta_k=n_k^{-1/3}$.  
In order to determine the exponent $b$, we employ $n_k=n$ and a steplike $\beta$, with $\beta=n^{-5/12}$
for $k<k^*$ and $\beta=n^{-1/3}$ for $k>k^*$, where $k^*$ is the level at which the two impedances
are comparable. Following West et al. \cite{ref5}, we find that $V_b$ is dominated by the large vessels, 
implying that the MMR allometric exponent is $b=6/7 \approx 0.86$.

So far we have assumed the elastic modulus to be independent of $k$. To take the variation of $E_k$ 
with level $k$ into account, we need a restriction that depends on $E_k$. First, we note 
that the circumferential tension in the tube wall is $\sigma ={\overline p}r_k/h_k$, where 
${\overline p}$ is the average arterial pressure. Since for large arteries ${\overline p}$ 
does not vary much, we take ${\overline p}$ constant. Then, we can study the behaviour 
of the elastic energy stored in the walls, namely 
$U=\int (\sigma^2/2E)~dV=({\overline p}^2/2)\sum_k (N_k r_k^3l_k/E_kh_k)$.
Using $r_k/h_k$ constant, we find $U\sim M$ in the case $E_k=E$. 
This proportionality between elastic energy and body mass can be understood as follows.
At each heart beat, the energy imparted to the blood by the heart \cite{ref5} is given by 
(cardiac output)/(heart rate) $\sim M$.
Part of this energy is used to push the blood towards the capillaries, while another part
is stored in the elastic arteries. The latter is needed to attenuate the pressure pulse,
so that blood flow in the capillaries is smooth, allowing the exchange of nutrients
and gases. Since this process is important irrespective of body size, hypothesis (vi) seems natural
and is indeed observed in the case $E_k=E$.

We now show that the same allometric exponent is found for MMR if we take the total elastic energy
proportional to $M$ instead of the mass of arteries. Assuming that the average arterial pressure has
approximately the same value for all large arteries, we have a new auxiliary function, namely
\begin{eqnarray}
F(r_k,l_k,E_kh_k)&=&\sum_{k=0}^N \frac {AE_k^{1/2}h_{k}^{1/2}} {N_kr_k^{5/2}}+\lambda_b\sum_{k=0}^NN_kr_k^2l_k+\nonumber \\
                        &+& \sum_{k=0}^N \lambda_kN_kl_k^3+\lambda_a\sum_{k=0}^N \frac{N_kr_k^3l_k}{E_kh_k}~~.
\end{eqnarray}
Note that we may treat $E_kh_k$ as a single variable and that we have incorporated all 
constants in the Lagrange multipliers. Requiring 
$\partial F/\partial l_j=\partial F/\partial r_j=\partial F/\partial (E_jh_j)=0$ gives us
$\beta_j=n_j^{-5/12}$ and $E_jh_j/r_j=constant$. In the capillary case, we again
take the average pressure as constant, since the major pressure drop occurs in the arterioles.
Using the impedance in Eq. (7) in the auxiliary function we obtain $\beta_j=n_j^{-1/3}$.
Therefore we have the same scenario as before and recover $b=6/7\approx 0.86$ for the MMR.

The elastic properties of the vessels are important under heavy exercise. 
As observed by Womersley \cite{ref21}, on the other hand, during basal metabolism the 
vessels can be approximated as almost rigid tubes because
(a) $V_b$ and ${\dot Q_0}$ have their minimum functional values, and
(b) heart rate is minimal, while the wavelength of the pressure pulse is maximal.
(For example, in humans, where the aorta length = $l_0\approx 40~cm$, $\lambda_{basal}\approx 4~m $ and 
$\lambda_{max}\approx 0,9~m$).
The volume flow, defined in Eq. (4), can be written in terms of $\partial p(z,t)/\partial z$ as
\begin{equation}
{\dot Q}=\frac{\pi r_k^2}{\omega \rho_0}\frac{i\partial p}{\partial z}~~.
\end{equation}
If $\lambda >> l_k$, we have that $\partial p/\partial z\approx\Delta p_k/l_k$, where
$\Delta p_k$ is the time-dependent pressure difference between the extremities of the tube. 
Then the impedance is
\begin{equation}
Z_k=\frac{\rho_0\omega l_k}{\pi r_k^2}~~.
\end{equation}
This is the same impedance found by Womersley \cite{ref21} for a rigid tube with an oscillatory pressure 
gradient when $r_k$ is large and viscous effects are negligible. Now the local impedance coincides with the hydrodynamic
resistance of the whole tube against $\Delta p$. 
To obtain $\beta$, we must then minimize the following auxiliary function
\begin{equation}
F(r_k,l_k)=\sum_{k=0}^N \frac {D l_k} {N_kr_k^2}+\lambda_b\sum_{k=0}^NN_kr_k^2l_k+\sum_{k=0}^N \lambda_kN_kl_k^3~~.
\end{equation}
The requirements $\partial F/\partial r_j= \partial F/\partial l_j= 0$
furnish $\beta_j=n_j^{-1/2}$, characteristic of the so-called area-preserving regime. 
In the limit of small $r_k$, viscosity dominantes so that the impedance is given by Poiseuille's formula
\begin{equation}
Z_k=\frac{8\eta l_k}{\pi r_k^4}~~.
\end{equation}
If we use this impedance in Eq. (11), we obtain $\beta_j=n_j^{-1/3}$. 
Therefore, assuming that $n_j=n$, Eq. (3) provides $b=3/4$ for basal metabolism.

In summary, recent investigations have established that maximum metabolic 
rate scales with an exponent larger than that of basal metabolism. 
This suggests that basal and maximum metabolic rates are governed by 
different mechanisms. We have shown, however, that both scaling regimes 
are predicted by a resource distribution network model \cite{ref5} if we take into 
account the distinct hydrodynamic and elastic properties of the circulatory 
system at each metabolic state. The model correctly predicts that MMR 
scales as $M^{6/7}$, maximum heart rate as $M^{-1/7}$ and muscular 
capillary density as $M^{-1/7}$. The theoretical and 
experimental basis for the 3/4-law for basal metabolism were recently 
questioned by Dodds et al. \cite{ref12}. However, Savage and coworkers extensively 
scrutinized the experimental data and confirmed that quarter-power scaling 
is pervasive in biology. Here we have shown that the 3/4-law is correctly 
predicted by the WBE model for mammals and birds when a different set of 
assumptions is adopted and the appropriate impedances are used.

 We are grateful to R. Dickman for a careful reading of the manuscript.
 We thank CNPq, CAPES and FAPEMIG, Brazil, for financial support.
 Electronic addresses: ${\ast}$ lbarbosa@fisica.ufmg.br,
 ${\dagger}$ gjmg@fisica.ufmg.br, ${\ddagger}$ jaff@fisica.ufmg.br.

\vfill

\end{document}